\documentclass[bibyear]{aa}

\usepackage[T1]{fontenc}

\usepackage[colorlinks=true,linkcolor=blue,citecolor=blue]{hyperref}

\usepackage{graphicx}
\usepackage{amsmath}
\usepackage{multicol,multirow}
\usepackage{tabularx}
\usepackage{booktabs} 
\usepackage{color,soul}
\usepackage{cleveref}
\usepackage{xspace} 
\usepackage{xcolor}
\usepackage[flushleft]{threeparttable}

\usepackage{txfonts}

\usepackage{natbib}
\bibpunct{(}{)}{;}{a}{}{,}

\newcommand{\chandra}{\textit{Chandra}\xspace}

\newcommand{\nustar}{\textit{NuSTAR}\xspace}

\newcommand{\nicer}{{NICER}\xspace }
\newcommand{\ixpe}{IXPE\xspace}

\newcommand{\kms}{km~s$^{-1}$\xspace}
\newcommand{\fluxcgs}{erg~s$^{-1}$~cm$^{-2}$\xspace}
\newcommand{\lumcgs}{erg~s$^{-1}$\xspace}
\newcommand{\pcm}{\,cm$^{-2}$\xspace}	% per cm-squared 
\newcommand{\pcmc}{\,cm$^{-3}$\xspace}
\newcommand{\source}{4U~1624--49\xspace}
\newcommand{\monk}{\textsc{monk}\xspace}

\def\tbabs{\texttt{TBabs}\xspace}

\def\diskbb{\texttt{diskbb}\xspace}
\def\thcomp{\texttt{thcomp}\xspace}
\def\bbodyrad{\texttt{bbodyrad}\xspace}

\def\polconst{\texttt{polconst}\xspace}
\def\relxill{\texttt{relxill}\xspace}
\def\relxillns{\texttt{relxillNS}\xspace}

\newcolumntype{C}{>{$\displaystyle}c<{$}}
\xspaceaddexceptions{+ - *}

\authorrunning{A. Gnarini et al.} 

\graphicspath{{./}{Figures/}}

\begin{document} 

\title{Constraining the geometry of the dipping atoll 4U~1624--49 with X-ray spectroscopy and polarimetry}

\author{Andrea Gnarini \inst{\ref{in:UniRoma3}}\fnmsep\thanks{E-mail: andrea.gnarini@uniroma3.it} 
\and M. Lynne Saade \inst{\ref{in:USRA},\ref{in:NASA.MSFC}}
\and Francesco Ursini \inst{\ref{in:UniRoma3}}
\and Stefano Bianchi \inst{\ref{in:UniRoma3}}
\and Fiamma Capitanio \inst{\ref{in:INAF-IAPS}}
\and Philip Kaaret \inst{\ref{in:NASA.MSFC}}
\and Giorgio Matt \inst{\ref{in:UniRoma3}}
\and Juri Poutanen \inst{\ref{in:UTU}}
\and Wenda Zhang \inst{\ref{in:NAO.Beijing}}
}

\institute{Dipartimento di Matematica e Fisica, Università degli Studi Roma Tre, via della Vasca Navale 84, I-00146 Roma, Italy \label{in:UniRoma3}
\and
Science \& Technology Institute, Universities Space Research Association, 320 Sparkman Drive, Huntsville, AL 35805, USA \label{in:USRA}
\and
NASA Marshall Space Flight Center, Huntsville, AL 35812, USA \label{in:NASA.MSFC}
\and
INAF -- Istituto di Astrofisica e Planetologia Spaziali, Via del Fosso del Cavaliere 100, 00133 Roma, Italy \label{in:INAF-IAPS}
\and 
Department of Physics and Astronomy, FI-20014 University of Turku, Finland \label{in:UTU} 
\and
National Astronomical Observatories, Chinese Academy of Sciences, A20 Datun Road, Beijing 100012, China \label{in:NAO.Beijing}
}

\date{Received XXX; accepted YYY}

% \abstract{}{}{}{}{} 
% 5 {} token are mandatory

\abstract{
    We present the spectro-polarimetric results obtained from simultaneous X-ray observations with \ixpe, \nustar, and \nicer of the dipping neutron star X-ray binary \source. This source is the most polarized Atoll source so far observed with \ixpe, with a polarization degree of $2.7\% \pm 0.9\%$ in the 2--8~keV band during the nondip phase and marginal evidence of an increasing trend with energy. The higher polarization degree compared to other Atolls can be explained by the high inclination of the system ($i \approx 60\degr$). The spectra are well described by the combination of soft thermal emission, a Comptonized component, and reflection of soft photons off the accretion disk. During the dips, the hydrogen column density of the highly ionized absorber increases while the ionization state decreases. The Comptonized radiation seems to be the dominant contribution to the polarized signal, with additional reflected photons that contribute significantly even though their fraction in the total flux is not high.
}


\maketitle

%%%%%%%%%%%%%%%%% INTRODUCTION %%%%%%%%%%%%%%%%%%

\section{Introduction}

The Imaging X-ray Polarimetry Explorer (\ixpe; \citealt{Weisskopf.etAl.2016,Weisskopf.2022}), successfully launched on 2021 December 9, has provided a new powerful tool for investigating the properties of several classes of X-ray astronomical sources. Accreting weakly magnetized neutron stars in low-mass X-ray binaries (NS-LMXBs) are among the brightest X-ray sources in the sky and they are therefore great targets for X-ray polarimetric observations. NS-LMXBs typically accrete matter via Roche-lobe overflow from a late main sequence or evolved white dwarf companion star. The presence of the NS surface interrupts the accretion flow, forming a boundary or spreading layer between the disk and the NS surface, extending also to high latitudes at higher accretion rates \citep{Inogamov.Sunyaev.1999,Popham.Sunyaev.2001,Suleimanov.Poutanen.2006}. 

Based on their tracks in the hard-color/soft-color diagrams (CCDs) or hardness--intensity diagrams (HIDs) and on their joint timing and spectral properties, NS-LMXBs are traditionally divided in Atoll- and $Z$-sources \citep{Hasinger.VanDerKlis.1989,VanDerKlis.1989}. The $Z$-sources are characterized by higher luminosities and accretion rates ($> 10^{38}$\,\lumcgs; \citealt{Migliari.Fender.2006}) and trace a wide $Z$-like pattern in the CCD \citep{VanDerKlis.2006}. Atolls cover a wide range in luminosities ($\sim 10^{36}-10^{38}$\,\lumcgs), and at high luminosities (or accretion rate), they show a well-defined ``curved banana'' branch in the CCDs, along which they move on timescales of hours to days, sometimes also exhibiting secular motion that does not affect the variability \citep{Migliari.Fender.2006}. According to the timing properties (low-frequency noise and kHz quasi-periodic oscillations; \citealt{VanDerKlis.etAl.1990}), the banana branch is further divided into the upper and lower banana. The harder regions of the CCD patterns are traced out at lower luminosities (or accretion rates), forming isolated patches (the island state; \citealt{Hasinger.VanDerKlis.1989}). In this state, the motion of the source in the CCD is much slower (from days to weeks) and does not often follow a well-defined track, as the secular and the branch motions have similar timescales. 

The X-ray emission of NS-LMXBs is traditionally modeled with two main spectral components: a thermal one dominating in the soft X-ray band (below 1 keV) and a harder component related to the inverse Compton scattering of the soft photons by a hot electron plasma. In the traditional Eastern model, the thermal emission is related to the accretion disk photons directly observed and modeled as a multitemperature black body \citep{Mitsuda.etAl.1984,Mitsuda.etAl.1989}, while the Comptonization of seed photons from the NS surface in a boundary or spreading layer between the NS surface and the inner edge of the disk produces the harder component \citep{Pringle.1977,Inogamov.Sunyaev.1999,Popham.Sunyaev.2001,Suleimanov.Poutanen.2006}. On the contrary, in the Western model, the soft thermal component is a single-temperature black body associated with the NS surface emission, while the hard component is due to Comptonization of the disk photons \citep{White.etAl.1988}. In addition to the primary continuum, the X-ray photons emitted by the NS surface or spreading layer may be reflected off the accretion disk (e.g., \citealt{DiSalvo.etAl.2009,Miller.etAl.2013,Mondal.etAl.2017,Mondal.etAl.2018,Ludlam.etAl.2019,Iaria.etAl.2020,Ludlam.etAl.2022,Ursini.etAl.2023}). The two emission models are largely degenerate spectroscopically and it is still a matter of debate as to which one is correct. Polarimetric measurements, combined with timing and spectral observations, might be the key to constraining the geometry of the system and its physical properties. 

\ixpe is a joint NASA and Italian Space Agency (ASI) mission equipped with three X-ray telescopes with polarization-sensitive imaging detectors \citep[gas-pixel detectors,][]{Costa.2001} operating in the 2--8 keV band. During its two-year prime mission, several NS-LMXBs were observed by \ixpe: four $Z$-class sources, namely Cyg~X-2 \citep{Farinelli.etAl.2023}, XTE~J1701--462 \citep{Cocchi.etAl.2023}, GX~5--1 \citep{Fabiani.etAl.2023}, and Sco~X-1 \citep{LaMonaca.etAl.2024}, and three Atoll-sources, namely GS~1826--238 \citep{Capitanio.etAl.2023}, GX~9+9 \citep{Chatterjee.etAl.2023,Ursini.etAl.2023}, and 4U~1820--303 \citep{DiMarco.etAl.2023.4U}. In addition, two peculiar NSs in X-ray binaries were observed by \ixpe: Cir~X-1 \citep{Rankin.etAl.2024} and GX~13+1 \citep{Bobrikova.etAl.2024}. Significant polarization was detected in most of the observed NS-LMXBs: as opposed to $Z$-sources, which can reach a polarization degree of up to $4-5\%$ along the horizontal branch, Atolls typically exhibit lower polarization ($\lesssim 2\%$). With the exception of GS~1826--238, for which only upper limits were derived, the polarization degree of all the other Atoll sources increases with energy, with a peculiar, rapid increase up to $9-10\%$ in the 7--8~keV band for 4U~1820--303. For most of the NS-LMXBs observed by \ixpe, the polarization signal is consistent with the result of the combination of Comptonization in a boundary or spreading layer plus the reflection of soft photons off the accretion disk. 

\source is a persistent Atoll NS-LMXB exhibiting periodic $\sim 6$\,h-long dips in its light curve occurring every $\sim 21$\,h \citep{Watson.etAl.1985}. During dips, the flux in the 1--10~keV energy band decreases by up to 75\% with respect to the persistent flux \citep{Iaria.etAl.2007}. These dips are thought to be related to the obscuration of the central source by a thickened region of the outer accretion disk, where the incoming matter from the companion star first enters the accretion disk \citep{White.Swank.1982} due to its high inclination ($i > 60^\circ$; \citealt{Frank.etAl.1987}). \source is classified as an Atoll source in the banana state, similarly to GX~9+9 but with a particularly high accretion rate of 0.5--0.8 $L_\text{Edd}$ \citep{Lommen.etAl.2005}. Unlike many other NS-LMXBs, no quasi-periodic oscillations have been detected \citep{Smale.etAl.2001,Lommen.etAl.2005}. When the source is not in the dip state, it may show flaring behavior up to 30\% above its normal 2--10~keV flux level \citep{Jones.Watson.1989}. The X-ray spectrum of \source is well described by a two-component emission model \citep{Church.Balucinska.1995,Balucinska.etAl.2000,Smale.etAl.2001,Iaria.etAl.2007,Xiang.etAl.2009}, with a single-temperature blackbody representing the NS emission plus a high-energy cut-off power law related to an extended accretion disk corona. The thermal component may be completely obscured during the deepest part of the dips, while the spectrum seems to arise mainly from the partially obscured Comptonized radiation. The obscuration of the Comptonization component is more gradual than that of the thermal component, which is typically very rapid. The \chandra High-Energy Transmission Grating Spectrometer (HETG) was used to observe Ca \textsc{xx} and Fe \textsc{xxv}/\textsc{xxvi} K$\alpha$ absorption lines in the spectrum. These could be produced in a photoionized absorber between the coronal radius and the outer edge of the accretion disk \citep{Iaria.etAl.2007,Xiang.etAl.2009}.

\ixpe observed \source on 2023 August 19, with a significant detection of polarization in the 2--8 keV band and a marginal hint of increasing polarization degree with energy \citep{Saade.etAl.2024}. Only an upper limit has been found during the central part of the X-ray dips. Using only \ixpe spectra, the contribution to the polarization of each spectral component was difficult to constrain. However, the increasing trend of the polarization with energy may suggest that the polarization probably arises from Comptonized radiation \citep{Saade.etAl.2024}. Following on from \cite{Saade.etAl.2024}, we report a new spectro-polarimetric analysis of the \ixpe observation, adding simultaneous observations performed with \nustar and \nicer. These two X-ray facilities provide wider spectral coverage, which is crucial for constraining the physical properties of the source, such as the temperature and optical depth of the Comptonizing region, and for studying the reflection component. In addition to the spectro-polarimetric analysis, we also performed numerical simulations with the Monte Carlo, general relativistic, radiative transfer code \textsc{monk} \citep{Zhang.etAl.2019,Gnarini.etAl.2022} in order to constrain the possible geometry of the system. 
The paper is structured as follows. In Sect. \ref{sec:Obs}, we describe the X-ray observations of \source performed with \ixpe, \nustar, and \nicer, along with the data reduction procedure. In Sect. \ref{sec:Spec}, we present our spectro-polarimetric analysis, while in Sect. \ref{sec:Results} we discuss the polarimetric results, with the addition of numerical simulations that we compare with results derived from observations. We finish in Sect. \ref{sec:Conclusions} with a brief summary and conclusions.

%%%%%%%%%%%%%%%%% OBSERVATIONS %%%%%%%%%%%%%%%%%%

\section{Observations and data reduction}\label{sec:Obs}

\begin{table}
\caption{Log of the observations of \source.}             
\label{table:Obs}      
\centering                                     
\begin{tabular}{l l c c}         
\toprule\toprule       
\noalign{\smallskip}
  Satellite & Obs. ID & Start Date (UTC) & Exp. (ks) \\   
\midrule     
\noalign{\smallskip}
  \ixpe & 02007301 & 2023-08-19 09:51:25 & 198.9 \\
  \nustar & 90901327002 & 2023-08-20 18:36:09 & 24.2 \\
  \nicer & 6203930102 & 2023-08-20 02:38:00 & 0.1 \\
  \nicer & 6203930103 & 2023-08-21 00:15:40 & 1.8 \\
  \nicer & 6203930104 & 2023-08-22 04:31:22 & 0.8 \\
  \nicer & 6203930105 & 2023-08-23 00:22:26 & 0.7 \\
\bottomrule                                            
\end{tabular}
\end{table}

% \begin{table}
% \caption{Starting and ending times of each dip in MJD.}             
% \label{table:Dip.Time}      
% \centering                                     
% \begin{tabular}{l l c}         
% \toprule\toprule       
% \noalign{\smallskip}
%   & Start Time (MJD) & End Time (MJD) \\   
% \midrule     
% \noalign{\smallskip}
%   Dip 1 & 60175.4296 & 60175.5913 \\
%   Dip 2 & 60176.2981 & 60176.4587 \\
%   Dip 3 & 60177.0870 & 60177.3226 \\
%   Dip 4 & 60177.9687 & 60178.2580 \\
%   Dip 5 & 60178.8698 & 60179.0598 \\
% \bottomrule                                       
% \end{tabular}
% \end{table}

\subsection{\ixpe}

\ixpe observed \source from 2023 August 19 09:51:25 UT to August 23 08:55:31 UT for a total exposure time of 198.9 ks (see Table \ref{table:Obs}). Time-resolved spectral and polarimetric data files were produced using the standard dedicated \textsc{ftools} procedure \citep[HEASoft v6.32;][]{Heasarc} and the latest available calibration files (CALDB v.20240125), considering the weighted analysis method presented in \cite{DiMarco.etAl.2022} with the parameter \texttt{stokes=Neff} in \textsc{xselect} (see \citealt{Baldini.etAl.2022} for more details). Source spectra and light curves have been extracted from a circular region of 120\arcsec\ radius centered on the source with radius computed in an iterative way that leads to maximization of the signal-to-noise ratio (S/N) in the 2--8 keV band, similar to the approach described in \cite{Piconcelli.etAl.2004}. The background was extracted from an annular region centered on the source with an inner and outer radius of 180\arcsec\ and 240\arcsec, respectively. Following the prescription by \cite{DiMarco.etAl.2023} for sources with $< 1$\,count\,s$^{-1}$\,arcmin$^{-2}$, the background rejection is applied and the residual background contribution is subtracted in the spectro-polarimetric analysis. However, we verified that the background subtraction does not significantly alter the results, especially the polarimetric measurements.

\begin{figure}
    \centering
    \includegraphics[width=0.475\textwidth]{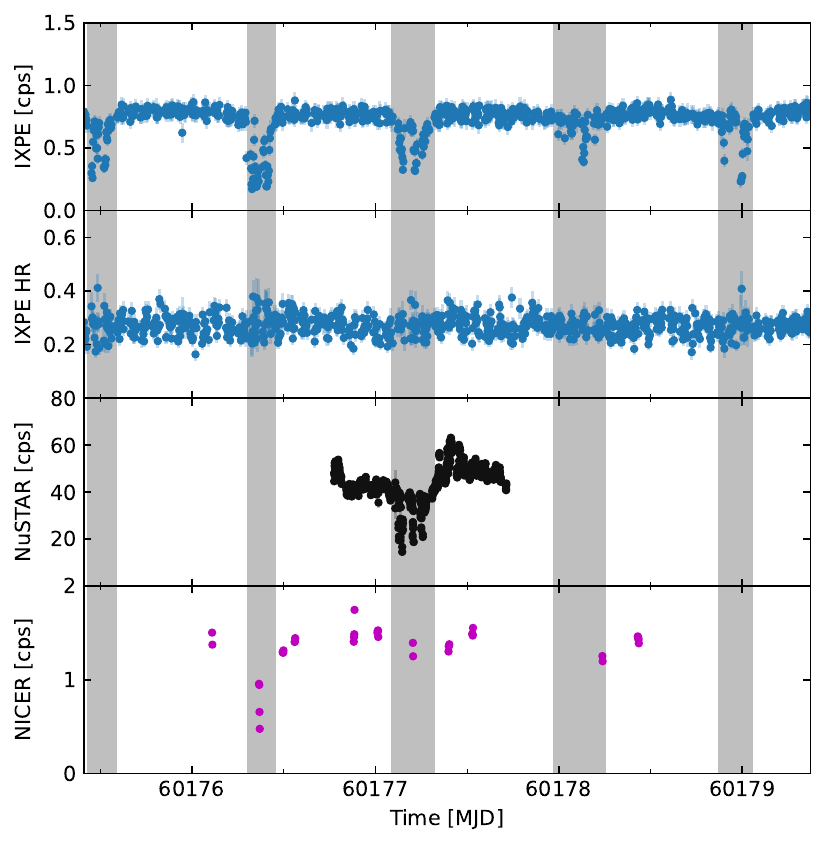}
    \caption{Light curves of \source during \ixpe, \nustar, and \nicer observations (in count\,s$^{-1}$). In the first two rows at the top, the \ixpe light curve (per detector) in the 2--8 keV energy band and the HR (5--8 keV/3--5 keV) are shown. In the third row, the \nustar 3--20 keV light curve (per detector) is represented. The \nicer 0.5--10 keV light curve (per detector) is plotted in the last row. The gray regions correspond to the dips. Each \ixpe bin corresponds to 480 s, while for \nustar and \nicer each bin is 120 s long.}
    \label{fig:LC.IXPE}
\end{figure}

The \ixpe light curve in the 2--8 keV energy range and the hardness ratio (HR: 5--8 keV/3--5 keV) are shown in Fig. \ref{fig:LC.IXPE}, with time bins of 120~s. During the \ixpe observation, five dips lasting between 15 and 25~ks are observed in the light curve, without any flaring activity. Therefore, we divided the observation into dip and nondip parts using the total count rate in the three detector units (DUs; \citealt{Saade.etAl.2024}): the source is in the nondipping state when the observed count rate for the 16~s bin light curve is $> 0.45$\,count\,s$^{-1}$, while the source is in a dipping state when the count rate is $< 0.45$\,count\,s$^{-1}$. The hardness ratio remains quite stable throughout the observation.

Some significant residuals remain in the \ixpe spectra, which are probably related to inter-calibration issues already observed in other sources. To account for these residuals, we apply the gain-shift correction to the \ixpe spectra with the \texttt{gain fit} command in \textsc{xspec}. The gain parameters of the $Q$ and $U$ spectra are linked to those of the $I$ spectra for each DU. The energy shift is derived using the relation $E' = E/\alpha - \beta$, where $\alpha$ is the gain slope and $\beta$ the offset.

\subsection{\nustar}

\begin{figure*}
    \centering
    \includegraphics[width=0.95\textwidth]{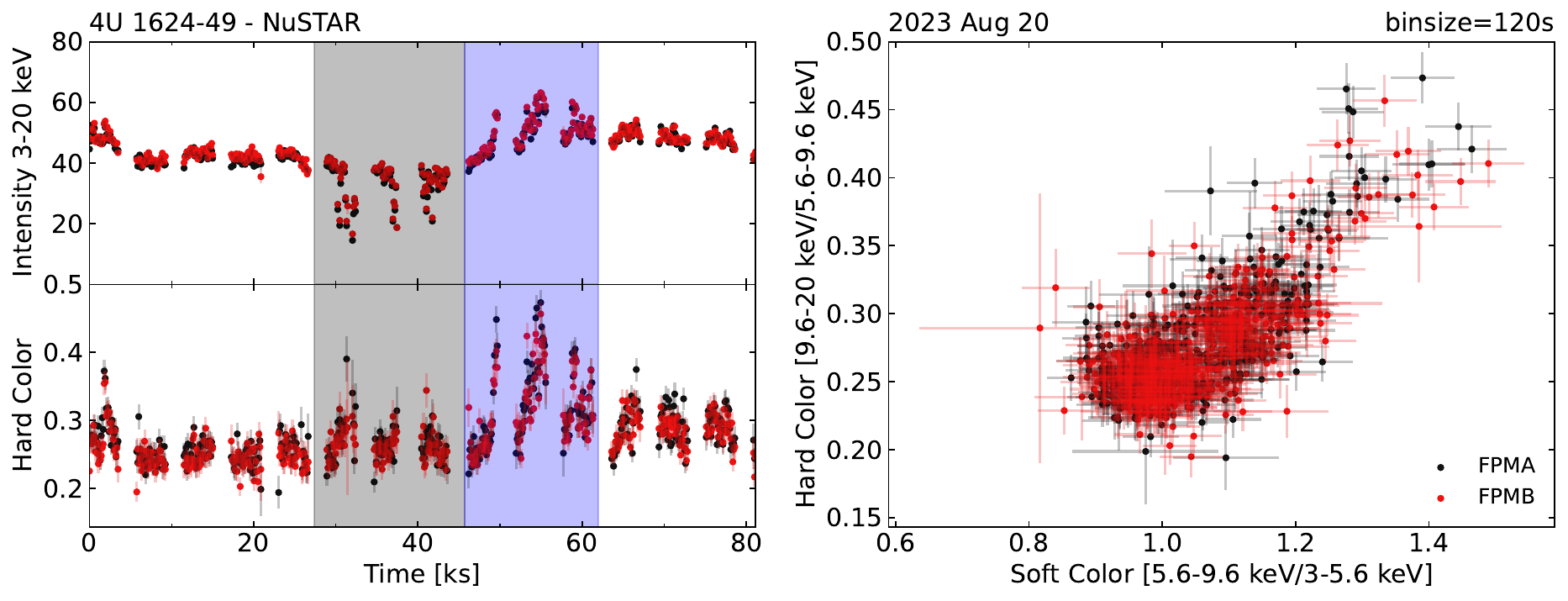}
    \caption{\nustar light curve (left panel) and CCD (right panel) of \source. The soft and hard colors are defined by Eqs.~\eqref{eq:Soft.Col} and \eqref{eq:Hard.Col}. The gray region corresponds to the dip. The blue region in the light curve highlights the motion of the source along the upper banana branch, corresponding to the points in the upper right part of the CCD diagram. Each bin corresponds to 120 s.}
    \label{fig:LC.CCD.Nustar}
\end{figure*}

The Nuclear Spectroscopic Telescope Array (\nustar; \citealt{Harrison.etAl.2013}) observed \source from 2023 August 20 18:36:09 UT to August 21 17:21:09 UT for a net exposure time of 24.2 ks (see Table \ref{table:Obs}). The unfiltered data were reduced using the latest calibration files (CALDB v.20240130) with the standard \texttt{nupipeline} and \texttt{nuproducts} tasks of the \nustar Data Analysis Software (\textsc{nustardas} v.2.1.2). For \nustar, the background is not negligible at all energies, and therefore we performed background subtraction. For both focal plane module (FPM) detectors, the regions used for spectra and light curve extraction were regions of  120\arcsec\ in radius  centered on the source and a background region of 60\arcsec\ in radius sufficiently far from the source. The source extraction radii were found with the same procedure of maximization of the S/N used for \ixpe data reduction \citep[see also][]{Piconcelli.etAl.2004,Marinucci.etAl.2022,Ursini.etAl.2023.CirGal}. Type-I X-ray bursts were not detected in any of the light curves. No quasi-periodic oscillations were detected with \nustar.

In Fig. \ref{fig:LC.CCD.Nustar}, we show the 3--20 keV \nustar light curve of \source, which spans the whole observation with 120 s time bins, along with the CCD. Soft and hard colors are defined as follows
\begin{equation}
    \label{eq:Soft.Col}
    \text{Soft Color} = \frac{\text{Count rate}~[5.6-9.6~\mathrm{keV}]}{\text{Count rate}~[3-5.6~\mathrm{keV}]}~,
\end{equation}
\begin{equation}
    \label{eq:Hard.Col}
    \text{Hard Color} = \frac{\text{Count rate}~[9.6-20~\mathrm{keV}]}{\text{Count rate}~[5.6-9.6~\mathrm{keV}]}~.
\end{equation}
The source remains in the banana state throughout the observation. During the \nustar exposure, only one dipping interval is observed between $\approx$ 30 and 45 ks (gray region in Fig. \ref{fig:LC.CCD.Nustar}), followed by a rapid increase in flux and hard color, corresponding to the motion of the source along the upper banana branch (blue region in Fig. \ref{fig:LC.CCD.Nustar}). The dip region is defined using the same time intervals derived from the \ixpe observation. We considered the data in the 3--20 keV range since the background starts dominating above 20 keV.

\subsection{\nicer}

The Neutron Star Interior Composition Explorer (\nicer; \citealt{Gendreau.etAl.2016}) observed \source multiple times from 2023 August 20 21:35:10 UT to August 23 05:26:20 UT for a total exposure time of 3.4 ks (see Table \ref{table:Obs}). The calibrated and cleaned files were processed using the standard \texttt{nicerl2} task of the \nicer Data Analysis Software (\textsc{nicerdas} v.12) together with the latest calibration files (CALDB v.20240206). The spectra and light curves were then obtained with the \texttt{nicerl3-spec} and \texttt{nicerl3-lc} commands, while the background was computed using the SCORPEON\footnote{\url{https://heasarc.gsfc.nasa.gov/docs/nicer/analysis_threads/scorpeon-overview/}} model. Due to their short duration, it is not possible to build a complete CCD or HID from the \nicer observations. The dip region is defined using the same time intervals derived from the \ixpe observation. No quasi-periodic oscillations were detected with \nicer. We noticed some relevant residuals in the \nicer energy spectrum below 2 keV \citep{Miller.etAl.2018,Strohmayer.etAl.2018}, likely related to spectral features not considered in the \nicer ancillary response file (ARF). To remove these features from the spectrum, we add a multiplicative absorption edge (\texttt{edge}). In particular, the threshold energy of the edge is found to be very close to the Al edge at 1.839 keV. The fit improves when fixing the energy to the same value as the Al edge (Tables \ref{table:Eastern.BestFit} and \ref{table:Western.Pol}).

%%%%%%%%%%%%%%%%% SPECTRA %%%%%%%%%%%%%%%%%%

\section{Spectro-polarimetric analysis}\label{sec:Spec}

We performed spectral and polarimetric analysis of the joint \ixpe+\nustar+\nicer observation using \textsc{xspec} (v.12.13.0; \citealt{Arnaud.1996}). As the polarization properties of the X-ray radiation coming from NS-LMXBs strongly depend on the geometry of the accreting system and the Comptonizing region, to test different emission models and configurations, the best-fits are obtained considering the following two spectral models:
\begin{description}
\item \texttt{TBabs*edge*CLOUDY*(diskbb + thcomp*bbodyrad + relxillNS)} ~,
\item
\item \texttt{TBabs*edge*CLOUDY*(thcomp*diskbb + bbodyrad + relxillNS)} ~,
\end{description}
which describe the Eastern- and Western-like models, respectively.

Energy-independent cross-calibration multiplicative factors were included for each \ixpe DU, each \nustar FPM, and the \nicer spectra. Interstellar absorption is modeled using \tbabs with \texttt{vern} cross-section \citep{Verner.etAl.1996} and \texttt{wilm} abundances \citep{Wilms.etAl.2000}. For the Eastern-like scenario, the model includes a multicolor disk blackbody (\diskbb; \citealt{Mitsuda.etAl.1984}) plus a harder Comptonized emission from the boundary or spreading layer \citep{Popham.Sunyaev.2001,Rev.etAl.2013} using the convolution model \thcomp \citep{Zdziarski.etAl.2020} applied to the \bbodyrad component. For the Western-like scenario, the convolution model \thcomp is applied to the multicolor disk blackbody \diskbb, while the NS emission is directly observed as a simple blackbody \bbodyrad. The covering factor $f$ of \thcomp corresponds to the fraction of Comptonized photons and is always $> 0.85$ for both models, with best-fit values equal to 0.99. To improve the fit statistic and to better constrain the other parameters of \thcomp, we fixed the covering factor to the best-fit value. Both \nicer and \nustar spectra exhibit absorption lines; therefore, an ionized absorber is added to the continuum model using \texttt{CLOUDY} \citep{Cloudy.2017}. The \texttt{CLOUDY} absorption table reproduces the absorption lines self-consistently through a slab with a constant density of $10^{12}$ \pcmc and a turbulence velocity of 500 \kms, illuminated by the unabsorbed intrinsic best-fit spectral energy distribution (SED) described below. A highly ionized plasma is required to model the absorption lines, with an ionization parameter of $\log \xi \approx 4.8$ and a H-column density of $N_{\rm Heq} \approx 10^{23}$ \pcm (during the nondip intervals).

The reflection component is required to remove the broadened iron line residuals between 6 and 7 keV in the \nicer and \nustar spectra. As no Compton hump is visible in the spectra, the reflection is produced by a softer illuminating spectrum with respect to the typical nonthermal (power-law) continuum considered in standard reflection models. In particular, we added \relxillns \citep{Garcia.etAl.2022} to both spectral models to take into account the reflected photons. The \relxill models are able to calculate the relativistic reflection from the innermost regions of the accretion disk \citep{Garcia.etAl.2014,Dauser.etAl.2014}. In particular, \relxillns assumes a single-temperature blackbody spectrum as the primary continuum that illuminates the disc, physically related to the NS surface or the spreading layer emission. This model assumes an incident illuminating spectrum at 45\degr\ on the surface of the disk \citep{Garcia.etAl.2022}. The emissivity index $q_\text{em} = 1.8$, the iron abundance $A_\text{Fe} = 1$, and the outer disk radius $R_\text{out} = 1000$ $R_\text{g}$ (where $R_\text{g}=G M/c^2$ is the gravitational radius) are fixed because the fit is relatively insensitive to these parameters (see also \citealt{Ludlam.etAl.2022}). For a standard NS, the spin period $P$ can be used to derive the dimensionless spin, adopting $a = 0.47/P(\mathrm{ms})$ \citep{Braje.etAl.2000}. Typical NS spin ranges between 2 and 5 ms \citep{Patruno.2017,DiSalvo.etAl.2023}; the dimensionless spin is then fixed at $a = 0.1$. We fixed the number density at $\log n_{\rm e}/\mathrm{cm}^{-3}= 16$, because the fit is not sensitive to this parameter given that its effect is only relevant at lower energies \citep{Ballantyne.2004,Garcia.etAl.2016}. The number density is consistent with the inner region of standard accretion disks (see \citealt{Garcia.etAl.2016,Ludlam.etAl.2022}). We also linked the temperature of the seed photons with the blackbody temperature of \bbodyrad. We tried to leave the inclination, the inner disk radius (in units of the ISCO), and the ionization parameter $\xi$ free to vary. However, the fit was not able to constrain the inner disk radius; therefore, it was fixed at 1.5 $R_\text{ISCO}$ for both models. The inclination of the system is found to be consistent within the errors between the two spectral models and with the presence of the dips in the light curves \citep{Frank.etAl.1987}. 

\begin{figure}
    \centering
    \includegraphics[width=0.475\textwidth]{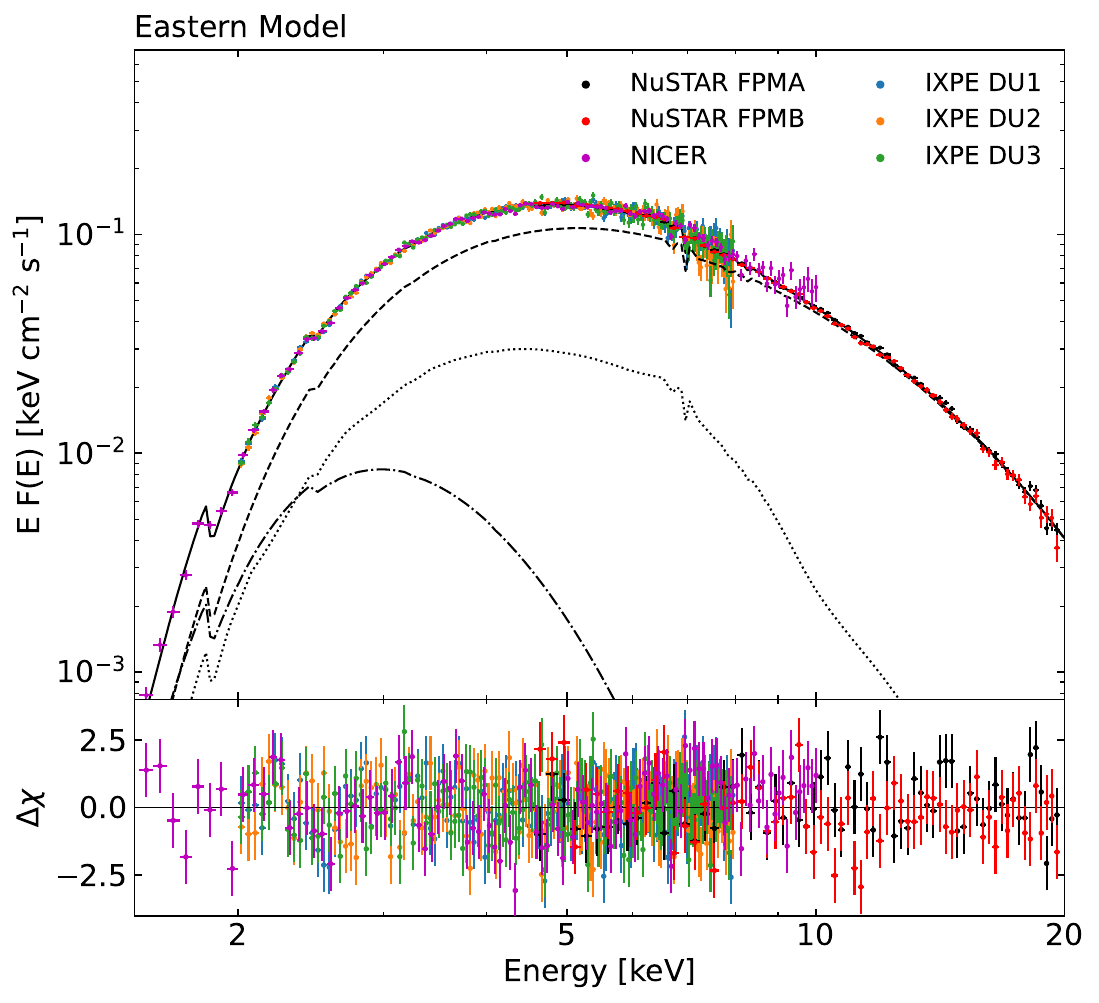}
    \includegraphics[width=0.465\textwidth]{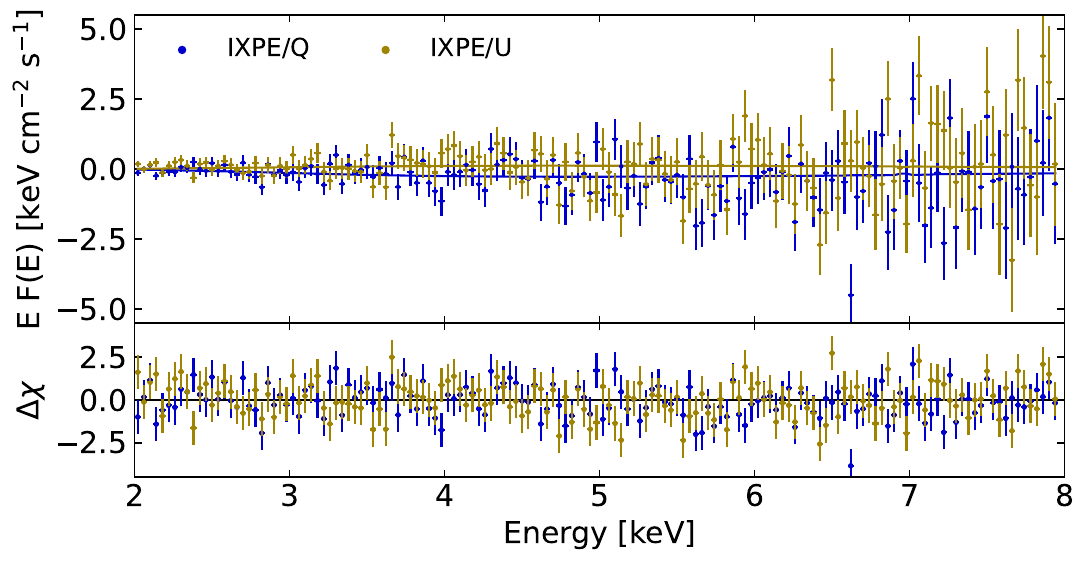}
    \caption{Deconvolved \ixpe (2--8 keV), \nustar (4.5--20 keV), and \nicer (1.5--10 keV) spectra (top panels) and \ixpe $Q$ and $U$ Stokes spectra (bottom panels), with the best-fit Eastern model and residuals in units of $\sigma$. The model includes \diskbb (dash-dotted lines), \thcomp*\bbodyrad (dashed lines), and \relxillns (dotted lines).}
    \label{fig:Spectra.Eastern}
\end{figure}

\begin{figure}
    \centering
    \includegraphics[width=0.475\textwidth]{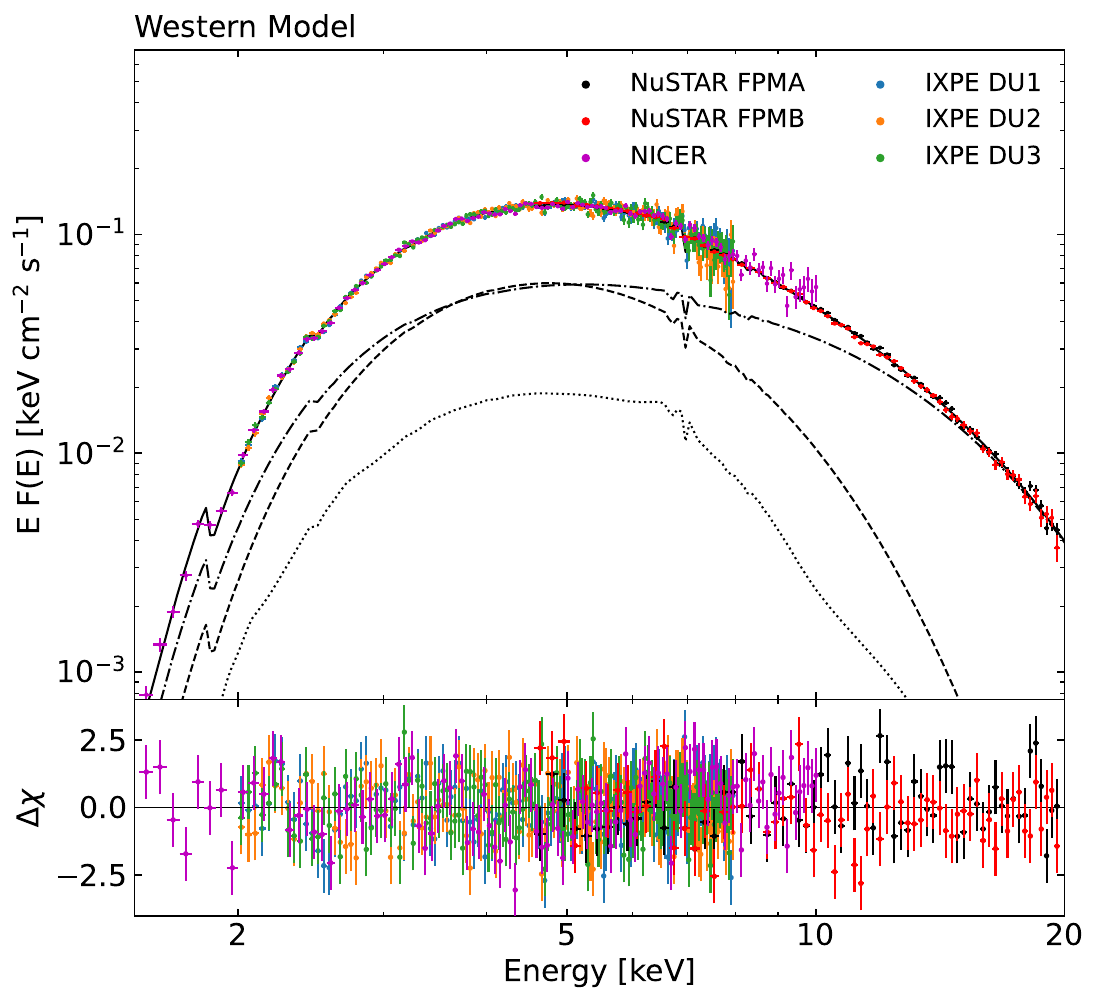}
    \includegraphics[width=0.465\textwidth]{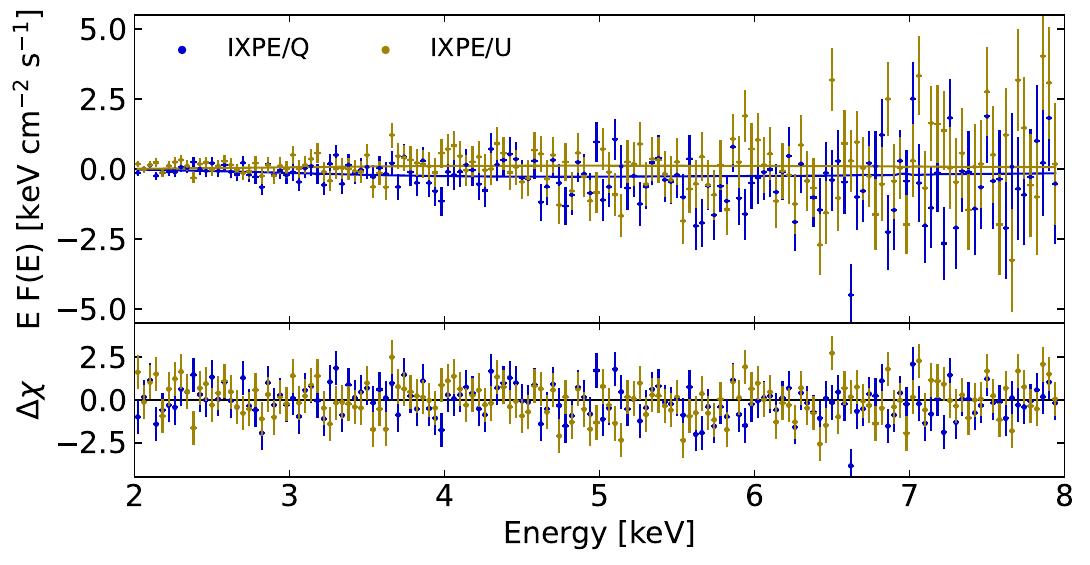}
    \caption{Deconvolved \ixpe (2--8 keV), \nustar (4.5--20 keV), and \nicer (1.5--10 keV) spectra (top panels) and \ixpe $Q$ and $U$ Stokes spectra (bottom panels), with the best-fit Western model and residuals in units of $\sigma$. The model includes \thcomp*\diskbb (dash-dotted lines), \bbodyrad (dashed lines), and \relxillns (dotted lines).}
    \label{fig:Spectra.Western}
\end{figure}

\renewcommand{\arraystretch}{1.25}

\begin{table}
\caption{Best-fit spectral parameters for the Eastern model obtained for \source with \ixpe+\nustar+\nicer spectra.} 
\label{table:Eastern.BestFit}      
\centering                                     
\begin{tabularx}{0.475\textwidth}{l l l l}        
\toprule\toprule         
\noalign{\smallskip}
Eastern model & Parameter & Nondip & Dip  \\   
\midrule     
\noalign{\smallskip}
\texttt{TBabs} & $N_{\rm H}$ ($10^{22}$\,cm$^{-2}$) & 10.6$^{+0.4}_{-0.6}$ & [10.6] \\
 % & $\tau_{\rm H}$ & [1.8] & [1.8] \\
\texttt{edge} & $E_{\rm c}$ (keV) & [1.839] & [1.839] \\
 & $D$ & 0.34$^{+0.07}_{-0.08}$ & 0.24$^{+0.16}_{-0.17}$ \\
\texttt{CLOUDY} & $\log \xi$ & 4.7$^{+0.2}_{-0.4}$ & 4.2$^{+0.1}_{-0.1}$ \\
 & $\log N_{\rm Heq}$ & 22.9$^{+0.3}_{-0.2}$ & 23.2$^{+0.1}_{-0.1}$ \\
\texttt{diskbb} & $kT_\text{in}$ (keV) & [0.6] & [0.6] \\
 & $R_\text{in}$ (km) & 20.6$^{+2.9}_{-3.5}$ & 18.6$^{+1.8}_{-1.8}$ \\
 \texttt{thcomp} & $\tau$ & 6.9$^{+0.8}_{-0.8}$ & [6.9] \\
 & $kT_{\rm e}$ (keV) & 3.2$^{+0.2}_{-0.2}$ & [3.2] \\
 % & $f$ & [1] & [1] \\
\texttt{bbodyrad} & $kT$ (keV) & 1.14$^{+0.02}_{-0.02}$ & [1.14] \\
 & $R_\text{bb}$ (km) & 13.3$^{+2.2}_{-2.4}$ & 11.8$^{+0.8}_{-0.7}$ \\
\texttt{relxillNS} & $q_\text{em}$ & [1.8] & [1.8] \\
 & $a$ & [0.1] & [0.1] \\
 & $i$ (deg) & 62$^{+12}_{-9}$ & [62] \\
 & $R_\text{in}$ (ISCO) & [1.5] &  [1.5] \\
 % & $R_\text{out}$ (ISCO) & [1000] & [1000] \\
 & $kT_\text{bb}$ (keV) & = $kT$ & = $kT$ \\
 & $\log \xi$ & 3.04$^{+0.07}_{-0.08}$ & [3.04] \\
 & $A_\text{Fe}$ & [1] & [1] \\
 & $\log n_{\rm e}$ & [16] & [16] \\
 & $N_{\rm r}$ ($10^{-4}$) & 10.6$^{+6.6}_{-2.9}$ & 10.7$^{+0.4}_{-0.4}$ \\
\midrule
  \multicolumn{2}{l}{$\chi^2/\text{d.o.f.}$} & 715.6/675 & 675.8/646 \\
  \multicolumn{2}{l}{$F_\text{Disk}/F_\text{Tot}$ (2--8 keV) \, \tablefootmark{a}} & 9.5\% & 9.3\% \\
  \multicolumn{2}{l}{$F_\text{Comp}/F_\text{Tot}$ (2--8 keV) \, \tablefootmark{b}} & 68.6\% & 64.1\% \\
  \multicolumn{2}{l}{$F_\text{Refl}/F_\text{Tot}$ (2--8 keV) \, \tablefootmark{c}} & 21.9\% & 26.6\% \\
  \multicolumn{2}{l}{$F_{\rm X}$ ($10^{-9}$ \fluxcgs) \, \tablefootmark{d}} & 1.45 & 1.17 \\
\bottomrule
\end{tabularx}
\tablefoot{
The errors are at the 90$\%$ confidence level for a single parameter. Parameters in square brackets were frozen during the fit. A 15 kpc distance is assumed to compute the inner radius $R_\text{in}$ and $R_\text{bb}$ from \diskbb and \bbodyrad normalizations. \\
\tablefoottext{a}{Percentage of disk photon flux in the 2--8 keV range.} \\
\tablefoottext{b}{Percentage of Comptonized photon flux in the 2--8 keV range.} \\
\tablefoottext{c}{Percentage of reflected photon flux in the 2--8 keV range.} \\
\tablefoottext{d}{Model flux in the energy range 0.1--20 keV.} \\
}
\end{table}

\begin{table}
\caption{Best-fit spectral parameters for the Western model obtained for \source with \ixpe+\nustar+\nicer spectra.} 
\label{table:Western.BestFit}      
\centering                                     
\begin{tabularx}{0.475\textwidth}{l l l l}        
\toprule\toprule         
\noalign{\smallskip}
Western model & Parameter & Nondip & Dip \\   
\midrule     
\noalign{\smallskip}
\texttt{TBabs} & $N_{\rm H}$ ($10^{22}$\,cm$^{-2}$) & 10.3$^{+0.2}_{-0.1}$ & [10.3] \\
 % & $\tau_{\rm H}$ & [1.8] & [1.8] \\
\texttt{edge} & $E_{\rm c}$ (keV) & [1.839] & [1.839] \\
 & $D$ & 0.31$^{+0.07}_{-0.07}$ & 0.28$^{+0.04}_{-0.05}$ \\
\texttt{CLOUDY} & $\log \xi$ & 4.7$^{+0.3}_{-0.3}$ & 4.2$^{+0.1}_{-0.1}$ \\
 & $\log N_{\rm Heq}$ & 22.8$^{+0.3}_{-0.2}$ & 23.2$^{+0.1}_{-0.1}$ \\
\texttt{thcomp} & $\tau$ & 15.8$^{+2.1}_{-2.2}$ & [15.8] \\
 & $kT_{\rm e}$ (keV) & 2.6$^{+0.1}_{-0.1}$ & [2.6] \\
 % & $f$ & [1] & [1] \\
\texttt{diskbb} & $kT_\text{in}$ (keV) & [0.6] & [0.6] \\
 & $R_\text{in}$ (km) & 18.9$^{+2.8}_{-2.9}$ & 16.9$^{+1.2}_{-1.2}$ \\
\texttt{bbodyrad} & $kT$ (keV) & 1.26$^{+0.02}_{-0.02}$ & [1.26] \\
 & $R_\text{bb}$ (km) & 8.4$^{+1.6}_{-1.7}$ & 7.2$^{+0.8}_{-0.7}$ \\
\texttt{relxillNS} & $q_\text{em}$ & [1.8] & [1.8] \\
 & $a$ & [0.1] & [0.1] \\
 & $i$ (deg) & 64$^{+18}_{-22}$ & [64] \\
 & $R_\text{in}$ (ISCO) & [1.5] & [1.5] \\
 % & $R_\text{out}$ (ISCO) & [1000] & [1000] \\
 & $kT_\text{bb}$ (keV) & = $kT$ & = $kT$ \\
 & $\log \xi$ & 2.93$^{+0.34}_{-0.33}$ & [2.93] \\
 & $A_\text{Fe}$ & [1] & [1] \\
 & $\log n_{\rm e}$ & [16] & [16] \\
 & $N_{\rm r}$ ($10^{-4}$) & 8.2$^{+4.5}_{-3.4}$ & 10.8$^{+1.9}_{-1.8}$ \\
\midrule
  \multicolumn{2}{l}{$\chi^2/\text{d.o.f.}$} & 716.5/675 & 683.5/646 \\
  \multicolumn{2}{l}{$F_\text{Comp}/F_\text{Tot}$ (2--8 keV) \, \tablefootmark{a}} & 46.6\% & 44.3\% \\
  \multicolumn{2}{l}{$F_\text{bb}/F_\text{Tot}$ (2--8 keV) \, \tablefootmark{b}} & 39.8\% & 33.6\% \\
  \multicolumn{2}{l}{$F_\text{Refl}/F_\text{Tot}$ (2--8 keV) \, \tablefootmark{c}} & 13.6\% & 22.1\% \\
  \multicolumn{2}{l}{$F_{\rm X}$ ($10^{-9}$ \fluxcgs) \, \tablefootmark{d}} & 1.45 & 1.17 \\
\bottomrule
\end{tabularx}
\tablefoot{
The errors are at the 90$\%$ confidence level for a single parameter. Parameters in square brackets were frozen during the fit. A 15 kpc distance is assumed to compute the inner radius $R_\text{in}$ and $R_\text{bb}$ from the \diskbb and \bbodyrad normalizations. \\
\tablefoottext{a}{Percentage of Comptonized photon flux in the 2--8 keV range.} \\
\tablefoottext{b}{Percentage of blackbody photon flux in the 2--8 keV range.} \\
\tablefoottext{c}{Percentage of reflected photon flux in the 2--8 keV range.} \\
\tablefoottext{d}{Model flux in the energy range 0.1--20 keV.} \\
}
\end{table}

Both models are able to provide excellent fits to the \ixpe+\nustar+\nicer spectra. The best-fitting parameters are reported in Tables \ref{table:Eastern.BestFit} and \ref{table:Western.BestFit}, respectively, for the Eastern-like and Western-like models, while the spectra are represented in Fig. \ref{fig:Spectra.Eastern} and \ref{fig:Spectra.Western}. The fit is not able to constrain the temperature at the inner disk radius, in particular for the Eastern-like scenario, because of the high absorption at lower energies, where the disk dominates. Therefore, we fixed $kT_\text{in}$ to 0.6 keV, which is consistent with the results of \cite{Iaria.etAl.2007}. The best-fitting parameters are quite typical for a NS-LMXB in the soft state, with the exception of the optical depth $\tau$ of \thcomp found with the Western-like model, which we find to be quite high. The ``apparent'' inner disk radius $R_\text{in}$ and the radius of the black body photon-emitting region $R_\text{bb}$ are derived from the normalization of \diskbb and \bbodyrad, assuming a 15 kpc distance \citep{Xiang.etAl.2009} and that all seed photons are Comptonized ($f=1$). The obtained results for $R_\text{in}$ and $R_\text{bb}$ are consistent within the errors between the two best-fit models, with the inner disk radius of $\sim 20 \sqrt{\cos i}$ km, where $i$ is the inclination of the system, and the black body radius is $\sim 10$ km, consistent with the boundary or spreading layer dimension or almost the entire NS surface.

The resulting slopes of the gain-shift correction are $0.983 \pm 0.005$, $0.954 \pm 0.004$, and $0.949 \pm 0.005$ keV$^{-1}$ for the three DUs, respectively, while their energy offsets are $0.064 \pm 0.019$, $0.130 \pm 0.021$, and $0.123 \pm 0.019$ keV for the Eastern-like scenario. For the Western model, these gain slopes are $0.982 \pm 0.006$, $0.953 \pm 0.006$, and $0.949 \pm 0.006$ keV$^{-1}$ with offsets of $0.067 \pm 0.022$, $0.135 \pm 0.023$, and $0.126 \pm 0.022$ keV, respectively. The gain-shift parameters are consistent between the two models, suggesting that they are indeed correcting for instrumental issues.

\subsection{Dust scattering halo}

During both persistent and dipping intervals, the galactic H-column density is relatively high ($\approx 10^{23}$\,cm$^{-2}$; see also \citealt{Parmar.etAl.2002,DiazTrigo.etAl.2006}) due to the presence of a dust-scattering halo \citep{Angelini.etAl.1997}. To model the presence of the halo, we followed the same approach as in \cite{DiazTrigo.etAl.2006}, considering the following two spectral models:
\begin{description}
\item \texttt{TBabs*edge*CLOUDY*e$^{-\tau_{\rm H}}$*(diskbb + thcomp*bbodyrad + relxillNS) + TBabs*edge*(1-e$^{-\tau_{\rm H}}$)*(diskbb + thcomp*bbodyrad + relxillNS)} ~,
\item
\item \texttt{TBabs*edge*CLOUDY*e$^{-\tau_{\rm H}}$*(bbodyrad + thcomp*diskbb + relxillNS) + TBabs*edge*(1-e$^{-\tau_{\rm H}}$)*(bbodyrad + thcomp*diskbb + relxillNS)} ~.
\end{description}
The first component of each model represents the contribution of the source reduced by a factor $e^{-\tau_{\rm H}}$, which corresponds to the scattering out of the line of sight, while the second component represents the scattering due to the halo. The parameter $\tau_{\rm H}$ is defined in terms of $\tau_1$, which is the optical depth at 1 keV as $\tau_1 E^{-2}$, where $E^{-2}$ is the theoretical energy dependence of the dust scattering cross-section. All the physical parameters of each spectral component are then tied between the source and the scattering halo emission. The fit is not able to provide an estimate of the optical depth $\tau_{\rm H}$, and therefore the optical depth is fixed at $\tau_{\rm H} = 1.8$, which is the best-fit value obtained by \citep{DiazTrigo.etAl.2006}.

Including the effect of the dust scattering halo, the fit is equivalent to those in Tables \ref{table:Eastern.BestFit} and \ref{table:Western.BestFit}. The best-fit values of most physical parameters do not significantly vary with the inclusion of the scattering halo in the spectral model; in particular, they are all consistent within the errors and the percentage difference of each parameter is less than 10\%. Only the normalizations of the spectral components vary slightly, with an increase of the reflection fraction and a small decrease in the thermal and Comptonized components. Also the polarimetric results are not strongly affected by the presence of the halo in the model: the total observed polarization in each energy band remains unchanged, while the polarization of each spectral component varies marginally according to the changes in its normalization. 

\subsection{Dip}

To check if the physical and polarimetric properties vary during the dips (i.e., the gray regions in Figs. \ref{fig:LC.IXPE} and \ref{fig:LC.CCD.Nustar}), we tried to perform the spectral analysis for the dip spectra as well. During the dips, no significant variations of the hard/soft colors are detected in the light curves. 

Since the spectral variations cannot be described as a simple increase in column density of the cold absorber \citep{Courvoisier.etAl.1986,Smale.etAl.1992}, there are different approaches to performing the spectral fitting during the dipping intervals: the first is to consider a two-component model with one absorbed component, whose column density increases significantly during the dips, and an unabsorbed component with decreasing normalization during the dips \citep{Parmar.etAl.1986}; the second approach would be to assume a blackbody (or multicolor disk blackbody) component partially obscured during the dip from an extended source modeled with a power law \citep{Church.Balucinska.1995,Balucinska.etAl.2000}. However, when including a highly ionized absorber in the model in order to take into account the absorption features detected in the spectra, the changes in the X-ray continuum can be explained by an increase in the column density and a decrease in the ionization state of the absorber \citep{DiazTrigo.etAl.2006}. 

Following the same approach, we performed the spectral fitting with the same model and physical parameters obtained for the nondip time interval, leaving only the column density, the ionization state of \texttt{CLOUDY,} and the normalization of each spectral component free to vary. The best-fitting parameters are reported in Tables \ref{table:Eastern.BestFit} and \ref{table:Western.BestFit} for the Eastern-like and Western-like models, respectively. For both models, the normalizations of the thermal and Comptonized components decrease, while the contribution of the reflection is more important. We obtained the same behavior for the highly ionized absorber as for other dipping NS-LMXBs \citep{DiazTrigo.etAl.2006}: the ionization parameter $\log \xi$ decreases during the dips, while there is an indication of the column density $N_{\rm H}$ increasing.

\subsection{Polarization}\label{sec:Polarization}

To estimate the polarization in the 2--8 keV band of \source, we applied the multiplicative model \polconst to the two best-fit models, with all the spectral parameters fixed to their best-fit values (Tables \ref{table:Eastern.BestFit} and \ref{table:Western.BestFit}). The \polconst model describes the energy-independent polarization degree ($\Pi$) and polarization angle ($\Psi$) for each component. As the two spectral models provide excellent fits to the spectra with very similar statistics, the polarimetric results applying \polconst to the entire model each time are almost identical: the polarization degree obtained during the nondip phase is $2.6\% \pm 0.9\%$ with a polarization angle of $78\degr \pm 10\degr$. Although the polarization is unconstrained in the 2--4 keV energy range at 99\% confidence level, the polarization increases up to $3.1\% \pm 1.3\%$ above 4 keV without any significant rotation in the polarization angle. Due to the low photon flux and the short duration of the dips, it is not possible to constrain the polarization during the dips, with only an upper limit of 3.3\% in the 2--8 keV energy band (at 99\% confidence level). This upper limit seems to exclude a strong increase in the polarization during the dips. All the polarimetric results are highly consistent within the errors with those obtained using the model-independent analysis with the \texttt{PCUBE} algorithm of \textsc{ixpeobssim} \citep{Baldini.etAl.2022}.

In order to derive the polarization properties of the X-ray emission of \source, we also performed the spectro-polarimetric fitting procedure by applying \polconst to each component of the best-fit models obtained from the broad-band spectral analysis:
\begin{description}
\item {\tt TBabs*edge*CLOUDY*(polconst}$^{(c)}${\tt *thcomp*bbodyrad + polconst}$^{(d)}${\tt *diskbb + polconst}$^{(r)}${\tt *relxillNS)}~,
\item
\item {\tt TBabs*edge*CLOUDY*(polconst}$^{(c)}${\tt *thcomp*diskbb + polconst}$^{(b)}${\tt *bbodyrad + polconst}$^{(r)}${\tt *relxillNS)}~,
\end{description}
respectively, for each best-fit model. All spectral parameters were fixed to their best-fit values (see Tables \ref{table:Eastern.BestFit} and \ref{table:Western.BestFit}), not including cross-calibration constants and leaving only the \polconst parameters free to vary.

%%%%%%%%%%%%%%%%% RESULTS %%%%%%%%%%%%%%%%%%

\begin{figure*}
    \centering
    \includegraphics[width=0.95\textwidth]{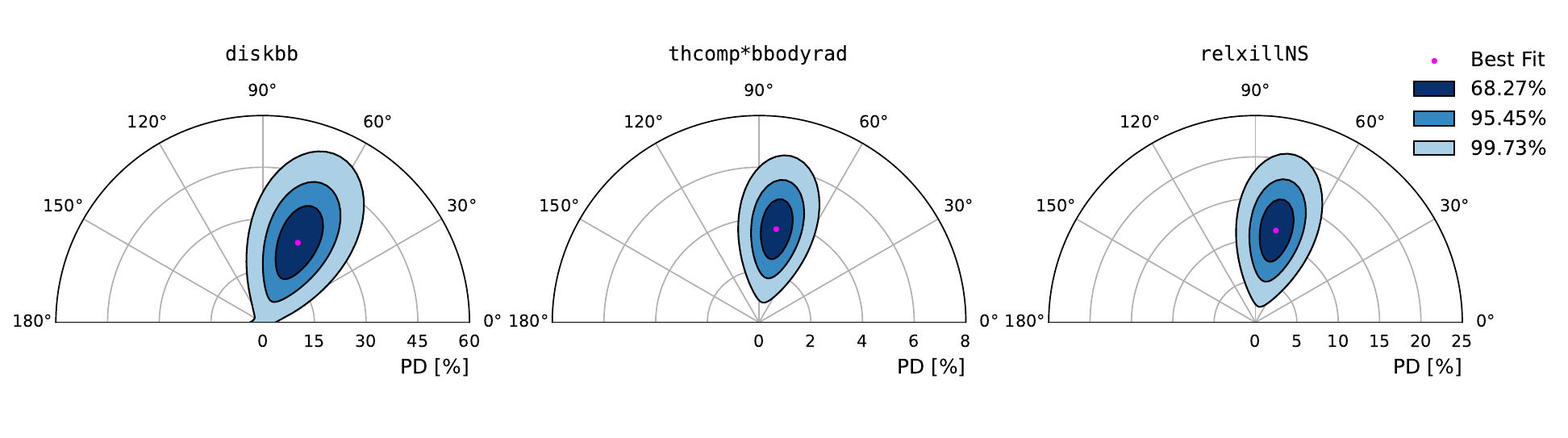}
    \caption{Polarization contour plots obtained with \textsc{xspec} assuming only one polarized component for the Eastern model: \diskbb (left panel), \thcomp*\bbodyrad (middle panel), and \relxillns (right panel). Contours correspond to the 68.27\%, 95.45\%, and 99.73\% confidence levels derived using a $\chi^2$ with two degrees of freedom. It is important to note the very different scales in the plots.}
    \label{fig:Contour.Eastern}
\end{figure*}

\section{Results and discussions}\label{sec:Results}

\begin{figure*}
    \centering
    \includegraphics[width=0.95\textwidth]{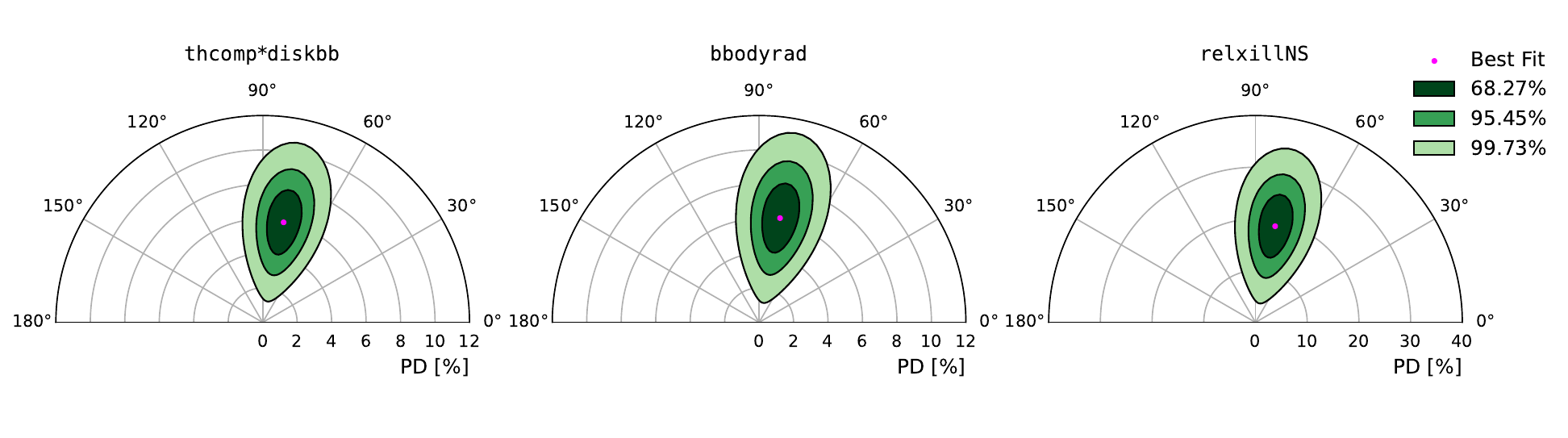}
    \caption{Polarization contour plots obtained with \textsc{xspec} assuming only one polarized component of the Western model: \thcomp*\diskbb (left panel), \bbodyrad (middle panel), and \relxillns (right panel). Contours correspond to the 68.27\%, 95.45\%, and 99.73\% confidence levels derived using a $\chi^2$ with two degrees of freedom. It is important to note the very different scales in the plots.}
    \label{fig:Contour.Western}
\end{figure*}

 The overlap of the energy range between \ixpe and \nustar+\nicer is expected to minimize systematic effects on the derived polarimetric parameters. However, it is difficult to constrain the polarization degree and angle for each spectral component because of the \ixpe band-pass. In particular, the Comptonized and reflected radiation are quite degenerate, peaking at similar energies and exhibiting similar spectral shapes. Some assumptions based on theoretical and observational expectations are required in order to estimate the contribution to the polarization of each spectral component. 

\subsection{Eastern model}

In order to estimate the polarization of each component, as a first test, we assumed that only one of the three spectral components is polarized, with the other two being unpolarized (Fig. \ref{fig:Contour.Eastern}). Assuming that only \diskbb is polarized, we find the fit to be acceptable ($\chi^2/\text{d.o.f.} = 1608.0/1586$, 897.1/894 for the subset of Stokes $Q$ and $U$ spectra), with $\Pi = 25\% \pm 12\%$, which is not well constrained and very high compared to the classical Chandrasekhar result for an inclination of $\approx 60\degr$ ($\sim 2-2.5\%$; \citealt{Chandrasekhar.1960}). The reason for such a high value of the polarization degree is that the contribution of disk photons in the 2--8 keV band is relatively low, as the disk is rather cold \citep{Iaria.etAl.2007}. Therefore, a high level of polarization is required to obtain the measured $\Pi$, because the disk emission will be depolarized by the other two (unpolarized) components. Indeed, the disk polarization is very poorly constrained given its small contribution in the \ixpe band. Moreover, the assumption of having only a polarized disk is not consistent with the observed polarization degree increasing with energy \citep{Saade.etAl.2024}: the opposite behavior is expected as the disk contribution to the total photon flux decreases with energy. 

The fit significantly improves considering the cases with polarized \thcomp*\bbodyrad or \relxillns ($\chi^2/\text{d.o.f.} = 1598.2/1586$, 882.8/894 for the subset of Stokes $Q$ and $U$ spectra). The model provides a good fit of the flux and Stokes $Q$ and $U$ spectra, without strong residuals. The polarization angle obtained in these cases is very similar and consistent within the errors: $80\degr \pm 10\degr$ is obtained for \thcomp*\bbodyrad only, while $78\degr \pm 11\degr$ is found for \relxillns alone. The reflection-dominated scenario requires a very large polarization degree of $11.4\% \pm 4.2\%$, compared to $3.7\% \pm 1.3\%$ obtained with only Comptonization. As the reflection is characterized by a polarization angle perpendicular to the disk plane \citep{Matt.1993,Poutanen.Svensson.1996,Schnittman.Krolik.2009}, its direction should be the same as that of the Comptonized component associated with the boundary or spreading layer \citep{Farinelli.etAl.2023,Ursini.etAl.2023}. 

By leaving the polarization degree of all spectral components, the polarization angle of the disk, and the Comptonized radiation free to vary, the contribution to the polarization of each component is difficult to constrain. As the disk contribution in the 2--8 keV range is small, we fixed its polarization at 2.5\%, corresponding to the classical results for a semi-infinite, plane-parallel, pure scattering atmosphere observed at $i \approx 60\degr$ \citep{Chandrasekhar.1960}. To estimate the polarization of the reflected radiation, we also set $\Pi$ of \thcomp*\bbodyrad to 1\%, a reasonable value for wedge- or spreading layer-like geometries for the inferred inclination \citep{Ursini.etAl.2023}. The resulting $\Pi$ of the reflected component is $8.8\% \pm 4.2\%$ with a polarization angle of $79\degr \pm 11\degr$, the same as Comptonization ($\chi^2/\text{d.o.f.} = 1598.4/1586$; Table \ref{table:Eastern.Pol}). By also leaving the polarization degree of the Comptonized radiation free to vary, only an upper limit of $13.9\%$ is obtained for \relxillns, while the \thcomp*\bbodyrad polarization is found to be $3.9\% \pm 2.6\%$, which is higher than expected for a typical quasi-spherically symmetric spreading layer-like geometry ($\chi^2/\text{d.o.f.} = 1596.9/1585$; Table \ref{table:Eastern.Pol}). We also tried to leave the polarization degree of the disk component free to vary, while $\Pi$ of the reflection was fixed at 10\% (see also \citealt{Matt.1993}), with that of Comptonization again fixed at 1\%. In this case, we obtain only an upper limit for the disk polarization of 19.2\% ($\chi^2/\text{d.o.f.} = 1598.2/1586$; Table \ref{table:Eastern.Pol}).

\subsection{Western model}

Similarly to the previous section, as a first case, we assumed that only one of the three spectral components is polarized, with the other two being unpolarized (Fig. \ref{fig:Contour.Western}). Assuming only \bbodyrad as the source of the polarized radiation, the obtained best-fit is acceptable ($\chi^2/\text{d.o.f.} = 1598.7/1586$, 885.1/894 for the subset of Stokes $Q$ and $U$ spectra) with $\Pi = 6.2\% \pm 2.2\%$ and $\Psi = 79\degr \pm 11\degr$. However, this assumption is not consistent with the observed $\Pi$ increasing with energy \citep{Saade.etAl.2024}. The fit slightly improves when considering only \thcomp*\diskbb or \relxillns ($\chi^2/\text{d.o.f.} = 1595.8/1586$, 881.0/894 for the subset of Stokes $Q$ and $U$ spectra) and the model is able to provide a good fit of both the flux and the Stokes $Q$ and $U$ spectra, without strong residuals. As the contribution of reflected photons in the 2--8 keV range is smaller than that of Comptonized radiation, the polarization degree obtained considering only \relxillns ($19\% \pm 7\%$) is significantly higher with respect to the case with only \thcomp*\diskbb ($5.9\% \pm 2.1\%$), while the two polarization angles are very similar and are consistent within the errors: $79\degr \pm 10\degr$ for only reflected components and $78\degr \pm 10\degr$ for only Comptonized radiation. 

As before, by leaving all the polarimetric parameters free to vary, the contribution to the polarization of each spectral component is impossible to constrain. As the blackbody component is associated with the NS surface emission, these photons should be unpolarized; therefore, the polarization of \bbodyrad is initially set to zero. We tried two different assumptions for the polarization of the reflection: since the reflected photons are expected to be polarized orthogonally to the accretion disk plane \citep{Matt.1993,Poutanen.Svensson.1996,Schnittman.Krolik.2009}, the polarization angle of \relxillns is initially set to be perpendicular to that of \thcomp*\diskbb. In order to estimate the contribution of the Comptonized disk radiation, we fixed $\Pi$ of \relxillns to 10\% \citep{Matt.1993}. In this case, we obtain a polarization degree of \thcomp*\diskbb of $9.0\% \pm 2.1\%$, while $\Psi$ is $78\degr \pm 10\degr$ ($\chi^2/\text{d.o.f.} = 1595.2/1586$; Table \ref{table:Western.Pol}). This value of $\Pi$ is significantly higher than the corresponding classic results for the same inclination \citep{Chandrasekhar.1960}. We then tried to leave the polarization degree or the angle of \relxillns free to vary in an attempt to obtain information about the polarization of reflection. However, it was not possible to find any constraints on the polarization of the reflected component for both cases with $\Pi$ or $\Psi$ of \thcomp*\diskbb free to vary. An alternative case is to set $\Psi$ of \relxillns to coincide with that of \thcomp*\diskbb. Under this assumption, we find $\Pi = 5.9\% \pm 2.2\%$ for \thcomp*\diskbb but only an upper limit of $16.6\%$ (at 90\% confidence level) for \relxillns ($\chi^2/\text{d.o.f.} = 1595.9/1585$). Also in this case, the resulting polarization degree for \thcomp*\diskbb is higher than the classical results \citep{Chandrasekhar.1960}.

We also tested a scenario in which the polarization of the NS is not zero, assuming that the \bbodyrad emission is associated to a boundary or spreading layer and with fixed $\Pi$ at 1\%. For a typical spreading layer with a vertical height greater than its radial extension ($H \gg \Delta R$), its polarization angle is expected to be orthogonal to the disk plane, as is the case for the reflected component. Similarly to the previous cases, it is not possible to find constraints for the polarization degree of reflection, and therefore we fixed it at 10\%. The resulting $\Pi$ of \thcomp*\diskbb is $9.9\% \pm 2.1\%$ with $\Psi$ of $78\degr \pm 10\degr$  ($\chi^2/\text{d.o.f.} = 1596.0/1586$; Table \ref{table:Western.Pol}). 

\begin{table}
\caption{Polarization degree and angle of each spectral component for different scenarios considering the Eastern-like model.} 
\label{table:Eastern.Pol}      
\centering                                     
\begin{tabularx}{0.475\textwidth} { 
  l
  >{\raggedright\arraybackslash}X 
  >{\raggedright\arraybackslash}X}        
\toprule\toprule         
\noalign{\smallskip}
 Component & $\Pi$ (\%) & $\Psi$ (deg) \\   
\midrule     
\noalign{\smallskip}
\diskbb & [2.5] & = $\Psi_\text{Comp} - 90\degr$ \\
\thcomp*\bbodyrad & [1] & $79 \pm 11$ \\
\relxillns & $8.8\pm 4.2$ & = $\Psi_\text{Comp}$ \\
\midrule
\diskbb & [2.5] & = $\Psi_\text{Comp} - 90\degr$ \\
\thcomp*\bbodyrad & $3.9 \pm 2.6$ & $80 \pm 10$ \\
\relxillns & $<13.9$ & = $\Psi_\text{Comp}$ \\
\midrule
\diskbb & $<19.2$ & = $\Psi_\text{Comp} - 90\degr$ \\
\thcomp*\bbodyrad & [1] & $80 \pm 11$ \\
\relxillns & [10] & = $\Psi_\text{Comp}$ \\
\bottomrule
\end{tabularx}
\tablefoot{
These spectro-polarimetric fits are statistically equivalent; i.e., the $\chi^2/\text{d.o.f.}$ is the same ($\chi^2/\text{d.o.f.} = 1.006$). The errors are at the 90$\%$ confidence level. Parameters in square brackets are frozen.
}
\end{table}

\subsection{Numerical simulations}

The polarization degree of the Comptonized and reflected radiation strongly depends on the geometry of the system. The expected X-ray polarization due to the Comptonizing region in NS-LMXBs was discussed by \cite{Gnarini.etAl.2022} based on detailed numerical simulations performed with the general relativistic Monte Carlo radiative transfer code \textsc{monk} \citep{Zhang.etAl.2019}. The polarization degree and angle depend on both the geometric configuration of the Comptonizing region and on the inclination of the source with respect to the line of sight \citep{Gnarini.etAl.2022,Capitanio.etAl.2023,Ursini.etAl.2023}. For \source, the polarization angle does not carry information because the orientation of the system is unknown.

Therefore, we consider two different geometries: an ellipsoidal shell around the NS equator to reproduce the boundary or spreading layer configuration, corresponding to the Eastern model, and a geometrically thin slab above the accretion disk, similar to the Western model. The elliptical shell is characterized by a semi-major axis that coincides with the inner edge of the accretion disk and extends up to a latitude of $45\degr$ on the NS surface. The optical depth is measured in the radial direction from the NS surface \citep{Popham.Sunyaev.2001}. The slab configuration is assumed to cover the entire accretion disk, starting from its inner edge $R_\text{in}$ and with vertical thickness $H \ll \Delta R$. The electron plasma for both the elliptical shell and the slab rotates with Keplerian velocity \citep{Zhang.etAl.2019,Gnarini.etAl.2022}. Differently to the spreading layer-like geometry, the optical depth for the slab geometry is defined as $\tau = n_{\rm e} \sigma_{\rm T} h$, where $n_{\rm e}$ is the electron number density, $\sigma_{\rm T}$ is the Thomson scattering cross section, and $h$ is the half-thickness of the slab (\citealt{Zhang.etAl.2019}; see also \citealt{Titarchuk.1994,Poutanen.Svensson.1996.2} for the same definition).

\begin{table}
\caption{Polarization degree and angle of each spectral component for different scenarios considering the Western-like model.} 
\label{table:Western.Pol}      
\centering                                     
\begin{tabularx}{0.475\textwidth} { 
  l 
  >{\raggedright\arraybackslash}X 
  >{\raggedright\arraybackslash}X}        
\toprule\toprule         
\noalign{\smallskip}
 Component & $\Pi$ (\%) & $\Psi$ (deg) \\   
\midrule     
\noalign{\smallskip}
\thcomp*\diskbb & $9.0 \pm 2.1$ & $78 \pm 10$ \\
\bbodyrad & [0] & $-$ \\
\relxillns & [10] & = $\Psi_\text{Comp} - 90\degr$ \\
\midrule
\thcomp*\diskbb & $5.9 \pm 2.2$ & $78 \pm 10$ \\
\bbodyrad & [0] & $-$ \\
\relxillns & $<16.6$ & = $\Psi_\text{Comp}$ \\
\midrule
\thcomp*\diskbb & $9.9 \pm 2.1$ & $78\pm 10$ \\
\bbodyrad & [1] & = $\Psi_\text{Comp} - 90\degr$ \\
\relxillns & [10] & = $\Psi_\text{Comp} - 90\degr$ \\
\bottomrule
\end{tabularx}
\tablefoot{
These spectro-polarimetric fits are statistically equivalent, i.e., the $\chi^2/\text{d.o.f.}$ is the same ($\chi^2/\text{d.o.f.} = 1.006$). The errors are at the 90$\%$ confidence level. Parameters in square brackets are frozen.
}
\end{table}

\begin{figure}
    \centering
    \includegraphics[width=0.475\textwidth]{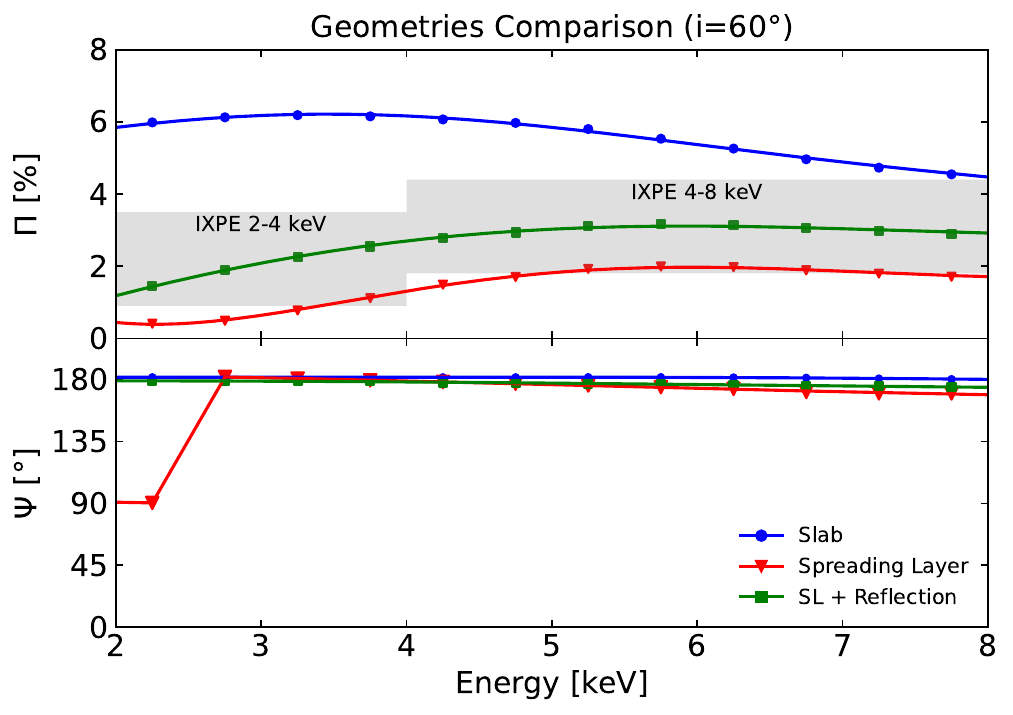}
    \includegraphics[width=0.235\textwidth]{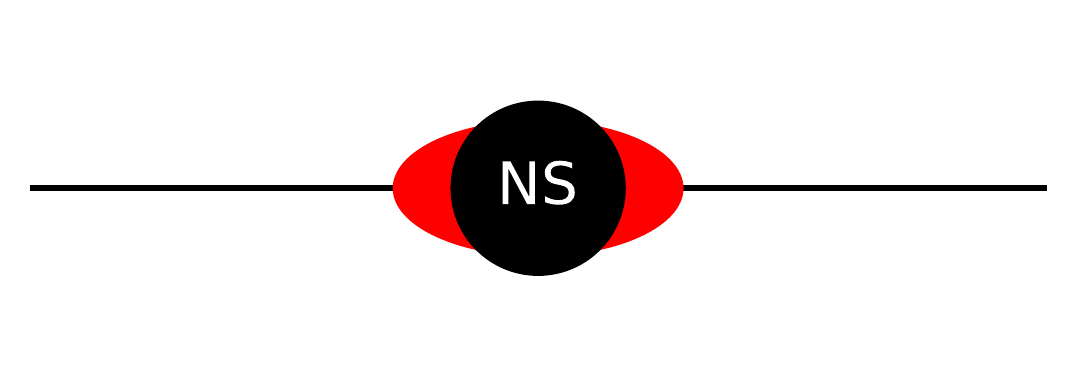}
    \includegraphics[width=0.235\textwidth]{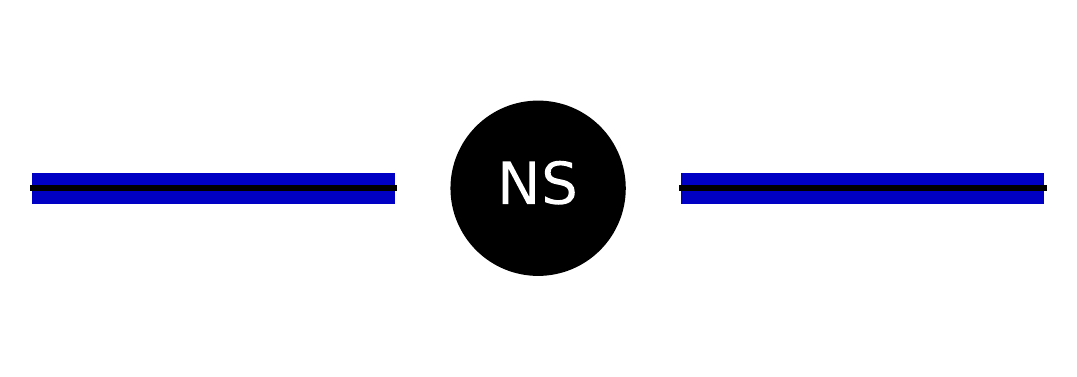}
    \caption{\textsc{monk} simulations of the $\Pi$ and $\Psi$ of the Comptonized+disk emission as a function of the energy in the 2--8 keV band for slab- and spreading layer-like configurations (blue and red respectively), along with a spreading layer case with the addition of the reflection component with $\text{P}_\text{refl} = 10\%$ and $\Psi_\text{refl} = 180$\degr\ (green). The inclination of the system is 60\degr. The gray regions correspond to \ixpe results at 90\% confidence level obtained by applying \texttt{polconst} to the two best-fit models (Sect. \ref{sec:Polarization}). The polarization angle in \textsc{monk} is measured from the projection of the rotation axis onto the sky plane.}
    \label{fig:MONK}
\end{figure}

Figure \ref{fig:MONK} shows the results for the total $\Pi$ and $\Psi$ obtained with \textsc{monk}. For the Eastern-like scenario, the polarized emission is the sum of the contribution of scattered NS photons and direct disk emission, which is considered to be polarized \citep{Chandrasekhar.1960}. Only a very small fraction of disk photons are expected to be scattered in the spreading layer and their contribution is negligible. The scattered NS photons are polarized orthogonally to the accretion disk plane, as the spreading layer is ---at the zero-order approximation--- perpendicular to the accretion disk plane, with a typical $\Pi$ of $\approx 1-2\%$ for $i \gtrsim 60\degr$ (see also \citealt{Ursini.etAl.2023}). The total polarization angle exhibits a rotation by $\approx 90\degr$: at lower energies, $\Psi$ is parallel to the accretion disk plane, where the direct disk photons dominate the emission, while at higher energies, the dominant scattered NS photons are polarized orthogonally. If the accretion rate is high enough to have the spreading layer covering the entire NS surface, similarly to a spherical shell geometry, the expected polarization degree is always $< 1\%$ for any inclination \citep{Gnarini.etAl.2022,Capitanio.etAl.2023}. As reflection is not yet included in \textsc{monk}, we can include its contribution in the total polarization with a simple Stokes vectorial analysis to estimate the total polarization. We can define three polarization pseudo-vectors for the spreading layer radiation, the direct disk emission, and the reflected photons:
\begin{equation}
\begin{aligned}
    \label{eq:Stokes.Vectors}
    & q_i = P_i f_i \cos 2 \Psi_i ~, \\
    & u_i = P_i f_i \sin 2 \Psi_i ~, 
\end{aligned}
\end{equation}
where $P_i$ and $\Psi_i$ are the polarization degree and angle of the three components, while $f_i$ is their relative contribution to the total photon flux in the 2--8 keV energy band, obtained from the best-fit model as reported in Table \ref{table:Eastern.BestFit}. If we assume a reflection component with $P_\text{refl} = 10\%$, $\Psi_\text{refl} = 0\degr$, and considering the results of \monk simulations for a spreading layer configuration ($P_\text{SL} = 1.5\%$, $\Psi_\text{SL} = 0\degr$) along with the direct disk contribution with $P_\text{disk} = 2.5\%$, $\Psi_\text{disk} = 90\degr$, which is consistent with the classical result for $i \approx 60\degr$ \citep{Chandrasekhar.1960}, the total expected polarization degree is
\begin{equation}
    \label{eq:Total.PD.Eastern}
    \text{P}_\text{tot} = \sqrt{q_\text{tot}^2 + u_\text{tot}^2} \approx 2.8\%
,\end{equation}
which is consistent with the observed $\Pi$. We applied a similar procedure to \monk simulations (Fig. \ref{fig:MONK}): by adding the contribution of reflection with $P_\text{refl} = 10\%$ and $\Psi_\text{refl} = 0\degr$, considering the fraction of reflected photons to the total 2--8 keV flux obtained from the best-fit, the resulting polarization degree for a spreading layer-like geometry is consistent with IXPE results in each considered energy band.

For the Western-like scenario, the polarized emission is the sum of the contribution of Comptonized disk photons and the NS photons scattered above the slab and ``reflected'' toward the observer. The direct emission from the NS surface is considered to be unpolarized. The scattered disk photons are polarized along the accretion disk but start to dominate the flux above 6--7 keV, with a slightly lower polarization degree than the classical result for a semi-infinite plane-parallel atmosphere above the disk \citep{Chandrasekhar.1960} due to special and general relativistic effects. The scattered NS photons are highly polarized orthogonally to the meridian plane, similarly to typical reflection of soft photons off the accretion disk by a centrally illuminating source \citep{Matt.1993}. This results in a relatively high polarization in the \ixpe energy band ($\gtrsim 5\%$ for $i > 60$\degr), which is only compatible with the observed polarization degree at higher energies. 

%%%%%%%%%%%%%%%%% CONCLUSIONS %%%%%%%%%%%%%%%%%%

\section{Conclusions}\label{sec:Conclusions}

The dipping source \source exhibits the highest polarization degree observed so far in the 2--8 keV band with \ixpe for Atoll sources. Simultaneous observations in the X-ray energy band were performed with \nustar and \nicer to constrain the physical parameters of the observed NS-LMXB. During the \ixpe observation, \source light curves display several dips, which are likely related to the obscuration of the central source by the outer region of the accretion disk \citep{White.Swank.1982,Frank.etAl.1987}. Because of the short duration and the low count rate during these dips, it was not possible to constrain the polarization in these periods, but only a 3.3\% upper limit is obtained. However, the variations of the spectra during the dip are consistent with an increase in column density with decreasing ionization state of the highly ionized absorber. The recent observation of the peculiar NS-LMXBs GX~13+1 showed a complex behavior of polarization during the dip \citep{Bobrikova.etAl.2024}. In particular, the polarization angle exhibits a rotation by $\approx 70^\circ$ between before and after the observed dip, with a corresponding variation of the polarization degree from $\approx 2\%$ to $\approx 5\%$. As opposed to the case of GX~13+1, no significant variations in the Stokes $Q$ and $U$ parameters are observed for \source during dips or immediately afterward. This seems to exclude the same increase in the polarization observed for GX~13+1 \citep{Bobrikova.etAl.2024}. Therefore, the physical mechanism responsible for the rotation of the polarization angle and the variation of the polarization degree for GX~13+1 seems not to be present for \source.  

The broad-band spectra of \source can be described with a soft thermal component plus a hard Comptonized emission, testing two different scenarios based on the ``classical'' spectral models \citep{Mitsuda.etAl.1984,White.etAl.1988}. The reflection of soft photons off the accretion disk is not negligible and must be considered in the spectro-polarimetric analysis. The main contribution to the polarization seems to come from the hard Comptonized emission for both scenarios, with the addition of the reflected component. In particular, comparing the \ixpe results with numerical simulations performed with \textsc{monk}, spreading-layer-like configurations for the Comptonizing region are able to reproduce the observed polarization if reflected photons are also considered, as they should be highly polarized \citep{Lapidus.Sunyaev.1985}. The slab configuration, on the other hand, is characterized by a higher polarization compared to the \ixpe results. Therefore, from the polarimetric observations, the Eastern-like scenario ---which is characterized by a spreading-layer geometry for the Comptonizing region--- seems to be favored over the Western-like model.

%--------------------------------------------------------------------

\begin{acknowledgements}
The Imaging X-ray Polarimetry Explorer (IXPE) is a joint US and Italian mission.  The US contribution is supported by the National Aeronautics and Space Administration (NASA) and led and managed by its Marshall Space Flight Center (MSFC), with industry partner Ball Aerospace (contract NNM15AA18C). AG, SB, FC, GM and FU acknowledge financial support by the Italian Space Agency (Agenzia Spaziale Italiana, ASI) through contract ASI-INAF-2022-19-HH.0.
%and by the Istituto Nazionale di Astrofisica (INAF). 
This research was also supported by the Istituto Nazionale di Astrofisica (INAF) grant 1.05.23.05.06: ``Spin and Geometry in accreting X-ray binaries: The first multi frequency spectro-polarimetric campaign''. 
JP thanks the Academy of Finland grant 333112 for support. 
WZ acknowledges the support by the Strategic Pioneer Program on Space Science, Chinese Academy of Sciences through grants XDA15310000, by the Strategic Priority Research Program of the Chinese Academy of Sciences, Grant No. XDB0550200, and by the National Natural Science Foundation of China (grant 12333004).

This research used data products provided by the IXPE Team (MSFC, SSDC, INAF, and INFN) and distributed with additional software tools by the High-Energy Astrophysics Science Archive Research Center (HEASARC), at NASA Goddard Space Flight Center (GSFC). This work was supported in part by NASA through the \nicer mission and the Astrophysics Explorers Program, together with the \nustar mission, a project led by the California Institute of Technology, managed by the Jet Propulsion Laboratory, and funded by the National Aeronautics and Space Administration. The \nustar Data Analysis Software (\textsc{NUSTARDAS}), jointly developed by the ASI Science Data Center (ASDC, Italy) and the California Institute of Technology (USA), has also been used in this project. This research has made use of data and/or software provided by the High Energy Astrophysics Science Archive Research Center (HEASARC), which is a service of the Astrophysics Science Division at NASA/GSFC.
\end{acknowledgements}


\bibliographystyle{aa} 
\bibliography{References.bib} 

\end{document}